\documentclass{birkjour_t2}
\usepackage[utf8]{inputenc}
\usepackage[english]{babel}
\usepackage{amsmath}
\usepackage{graphicx}

\begin{document}
\title[Application of conformal mapping technique to problems of DC distribution]{Application of conformal mapping technique to problems of direct current distribution in thin film wires bent at arbitrary angle}

\author{T.~N.~Gerasimenko and P.~A.~Polyakov}

\begin{abstract}
Current distributions in thin film wires bent at different angles were investigated with a conformal mapping method. The technique of angles rounding using three parameters was suggested. This technique enabled to obtain a smooth line having a similar form with an arc of a circle and to avoid infinite current density in the angle. The dependency between current density and the radius of rounding was examined.
\end{abstract}

\maketitle


\section*{Introduction}
Current distributions in flat conductors are problems of interest because of printed circuit boards and on-chip devices development \cite{Minimisation, Gijs, Panhorst}. Commonly these problems are solved numerically \cite{Petersen} or using near-field measurements \cite{nearfield1, nearfield2}. In the event if there are various angles in conductors the numerical solutions have bad convergence near the angles because formally the current density in the angle is infinite. In real conductors there are not such problems because there are not absolutely sharp angles and a small rounding always exists. It is difficult to take this rounding into account numerically therefore analytical methods are required. Such methods were investigated by P. M. Hall \cite{Hall_patterns, Hall_JAP, Hall_rectangles} and L. N. Trefethen \cite{Trimming} but only the analytical solution for angle of 90$^{\circ}$ with rounding was found \cite{Hall_JAP}.
In this paper we have considered current distributions in thin film wires bent at different angles with rounding and suggested the technique of estimating an optimal radius of corner rounding in such wires.

\section{The conformal mapping for a conductor bent at arbitrary angle}

\begin{figure}[!t]
\centering
\includegraphics[width=2.0in]{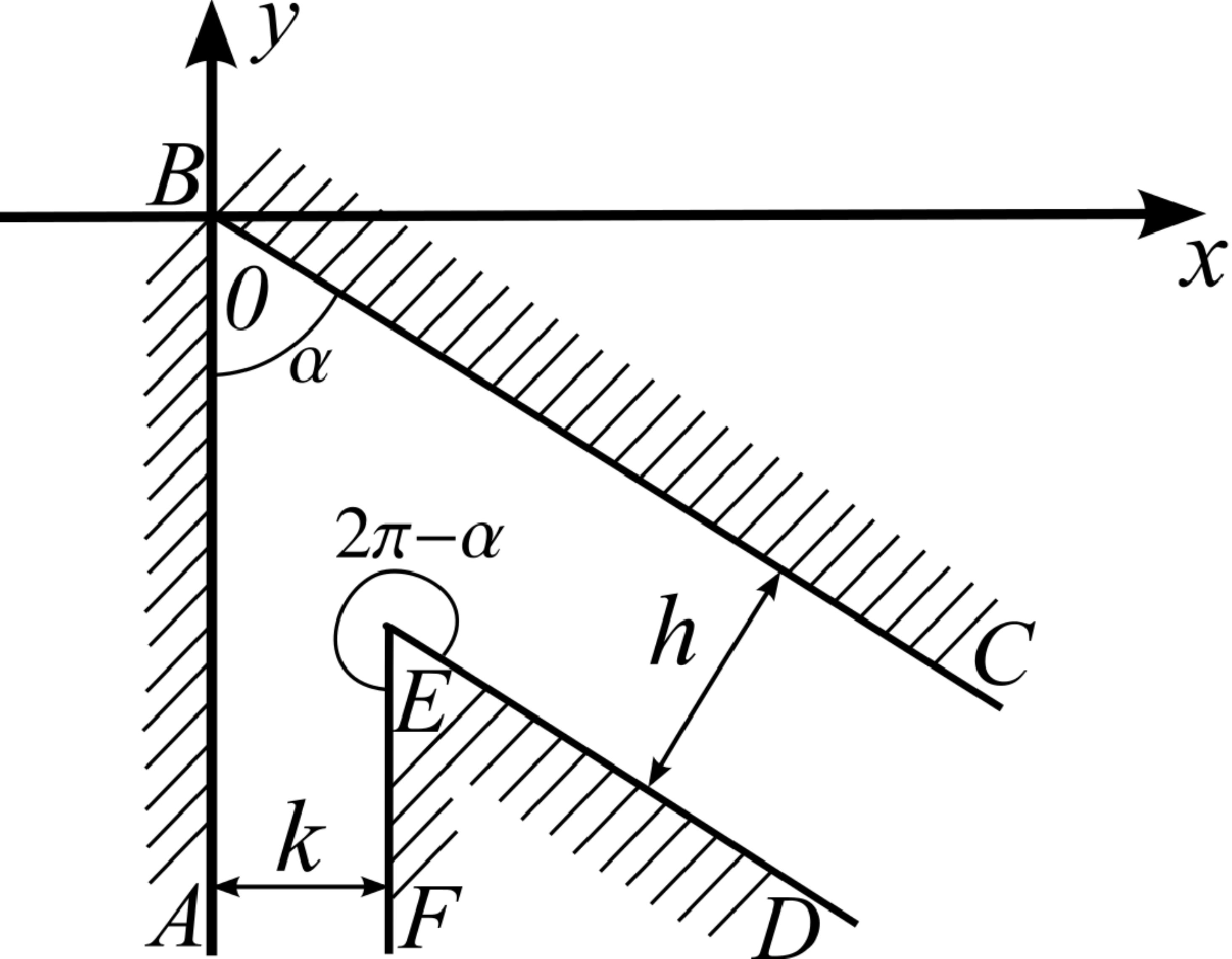}
\caption{The geometry of considered conductor.}
\label{post_zadachi}
\end{figure}

The geometry of a considered conductor is depicted in figure \ref{post_zadachi}. $\alpha$ is an arbitrary angle. Only the case of $h>k$ have been investigated  (the other may be examined in a similar way). We have considered a direct current. This fact has enabled us to reduce Maxwell's equations to the Laplace equation for a scalar potential
\begin{equation}
\Delta\varphi=0.
\end{equation}

This equation was solved for a complex potential $W(z)=U(z)+iV(z)$ using the conformal mappings method; $V(z)$ is a scalar potential and $U(z)$ is a stream function. These two functions are bound with Cauchy--Riemann conditions therefore a boundary conditions can be written only for one of them. The boundary conditions requires the absence of a current flow through lateral boundaries of the conductor so the resulting boundary value problem is the following \cite{Gerasimenko}

\begin{displaymath}
\left\{
\begin{array}{l}
\dfrac{\partial^2 W}{\partial x^2}+\dfrac{\partial^2 W}{\partial y^2}=0\\
\left. U(x,y)\right|_{(x,y)\in\Omega_1}=U_1=\mathrm{const}\\
\left. U(x,y)\right|_{(x,y)\in\Omega_2}=U_2=\mathrm{const}\\
\end{array}
\right.
\end{displaymath}
$\Omega_1$ and $\Omega_2$ are the upper and lower boundaries of the conductor respectively.

To solve this problem we have obtained first the solution of Laplace equation in the upper complex half-plain with the boundary conditions corresponding a point charge at the origin. Than we have mapped this solution onto the considered domain.

The required mapping corresponds the Schwarz--Cristoffel transformation of the following type
\begin{equation}
z=C\int\frac{1}{z_1}\left(\frac{z_1-1}{z_1+a}\right)^{1-\beta}dz_1
\label{Schwarz-Cristoffel}
\end{equation}
$\beta=\alpha/\pi$. Undefined constants $C$ and $a$ have been determined from the conditions of boundary conformity and having the following view
\begin{gather*}
a=\left(\frac{h}{k}\right)^{\tfrac{1}{1-\beta}}\\
C=\frac{h}{\pi}(\sin\alpha-i\cos\alpha)
\end{gather*}
where $h$, $k$ and $\alpha$ are depicted in figure~\ref{post_zadachi}.

The expression (\ref{Schwarz-Cristoffel}) is representable in terms of elementary functions if $1-\beta=P/Q$ where $0<P< Q$; $P$ and $Q$ are integers \cite{ConfDict}. From a physical point of view this condition does not impose any restrictions on an angle because any irrational number can be approximated by the rational one with arbitrary high accuracy. Thus, (\ref{Schwarz-Cristoffel}) can be reduced to
\begin{equation}
z=-(1+a)QC\int\frac{t^{Q-1}}{t^P(t^{Q}+a)(t^Q-1)}dt
\label{Int}
\end{equation}
where
\begin{displaymath}
t=\left(\frac{z_1+a}{z_1-1}\right)^{\tfrac{1}{Q}}.
\end{displaymath}

\begin{figure}
\centering
\includegraphics[width=3.4in]{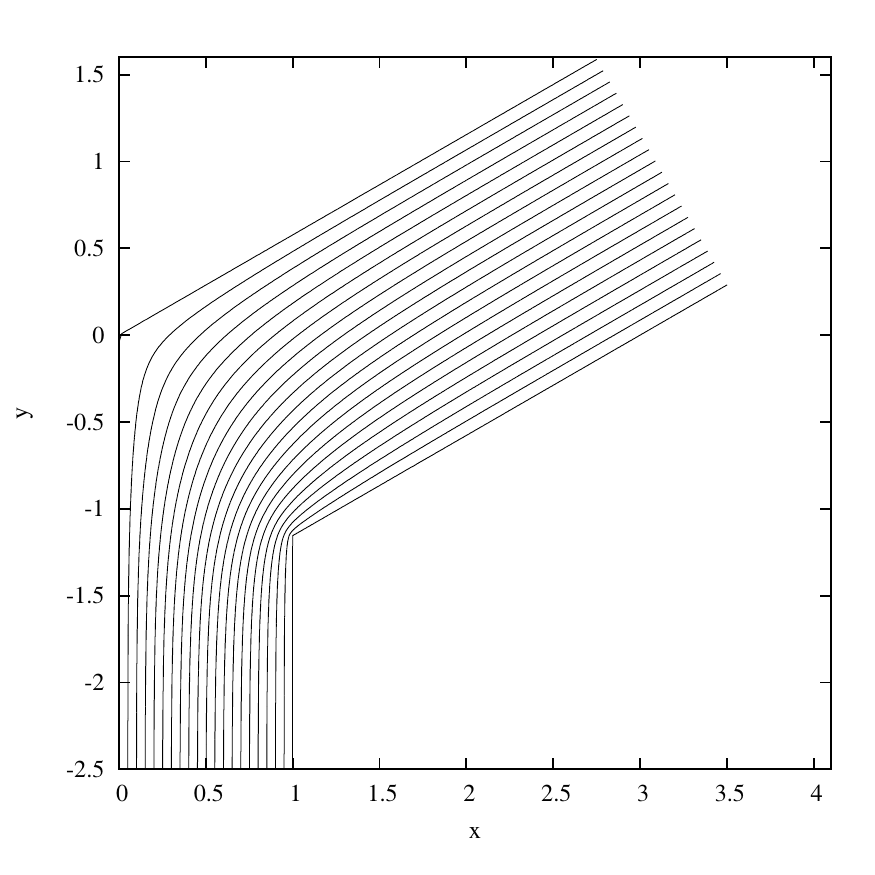}
\caption{Distribution of a current density for  $\alpha=120^{\circ}$ and $h/k=1.5$.}
\label{120}
\end{figure}

We have integrated the expression (\ref{Int}) for the angles of 60$^{\circ}$, 30$^{\circ}$, 120$^{\circ}$ and obtained the following results:

\begin{enumerate}
\item $\alpha=60^{\circ}$:
\begin{multline}
z=C\left\{\frac{1}{b^2}\ln(t+b)
-\frac{1}{2b^2}\ln(t^2-bt+b^2)
-\ln(t-1)
+\frac{1}{2}\ln(t^2+t+1)\right.
\\
\left.+\frac{i\sqrt{3}}{2b^2}\ln\left(\frac{-2t+b-i\sqrt{3}b}{2t-b-i\sqrt{3}b}\right)
+\frac{i\sqrt{3}}{2}\ln\left(\frac{-2t-1-i\sqrt{3}}{2t+1-i\sqrt{3}}\right)\right\}
+C_1
\label{Expr60}
\end{multline}
where
\begin{gather*}
C=\frac{h}{2\pi}(\sqrt{3}-i)\\
C_1=-\frac{h+k}{4}-i\frac{\sqrt{3}}{12}(7k-h)\\
t=\left(\frac{z_1+b^3}{z_1-1}\right)^{\tfrac{1}{3}}\\
b=\sqrt{\frac{h}{k}}
\end{gather*}

\begin{figure}
\centering
\includegraphics[width=3.4in]{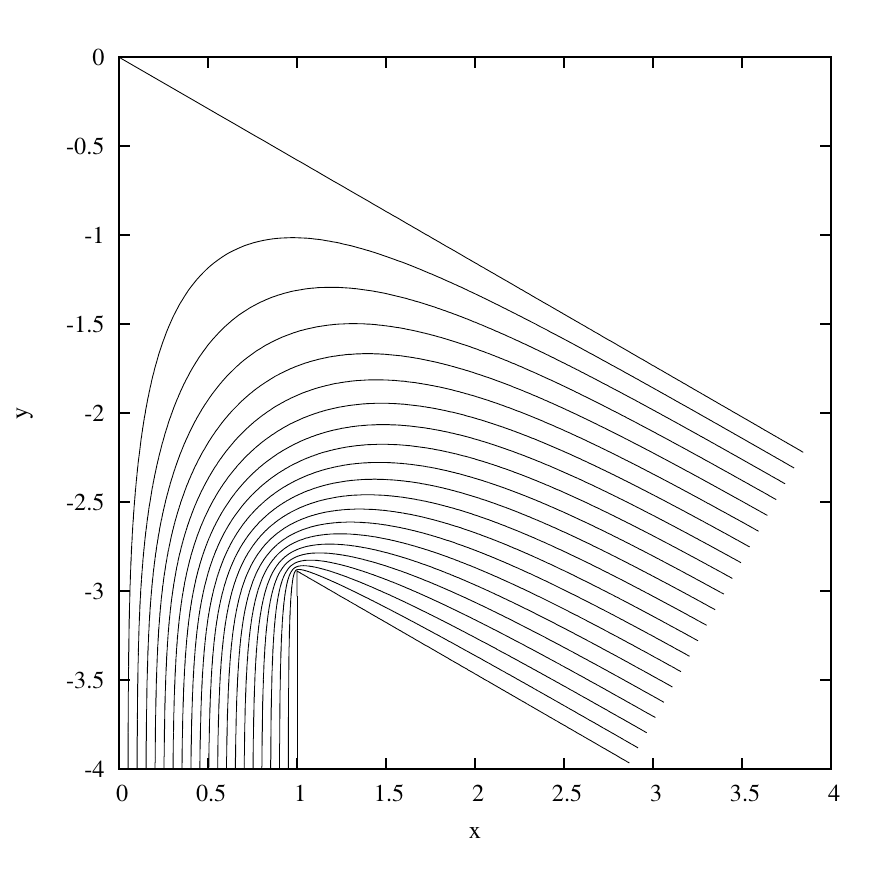}
\caption{Distribution of a current density for  $\alpha=60^{\circ}$ and $h/k=2$.}
\label{60}
\end{figure}

\item  $\alpha=30^{\circ}$:
\begin{multline}
z=C\left\{
\frac{i}{b^5}\ln\left(\frac{t+ib}{-t+ib}\right)
+\frac{\sqrt{3}}{2b^5}\ln\left(\frac{t^2+\sqrt{3}bt+b^2}{t^2-\sqrt{3}bt+b^2}\right)
+\frac{i\sqrt{3}}{2}\ln\left(\frac{2t+1+i\sqrt{3}}{-2t-1+i\sqrt{3}}\right)\right.
\\
+\frac{i\sqrt{3}}{2}\ln\left(\frac{2t-1+i\sqrt{3}}{-2t+1+i\sqrt{3}}\right)
+\frac{i}{2b^5}\ln\left(\frac{2t+\sqrt{3}b+ib}{-2t-\sqrt{3}b+ib}\right)
+\frac{i}{2b^5}\ln\left(\frac{2t-\sqrt{3}b+ib}{-2t+\sqrt{3}b+ib}\right)
\\
+\left.\ln\left(\frac{t+1}{t-1}\right)
+\frac{1}{2}\ln\left(\frac{t^2+t+1}{t^2-t+1}\right)\right\}+C_1
\label{Expr30}
\end{multline}
where
\begin{gather*}
C=\frac{h}{2\pi}(1-i\sqrt{3})\\
C_1=k-h\frac{\sqrt{3}}{2}-i\left(\frac{h}{2}+k\sqrt{3}\right)\\
t=\left(\frac{z_1+b^6}{z_1-1}\right)^{\tfrac{1}{6}}\\
b=\left(\frac{h}{k}\right)^{\tfrac{1}{5}}
\end{gather*}

\begin{figure}
\centering
\includegraphics[width=3.4in]{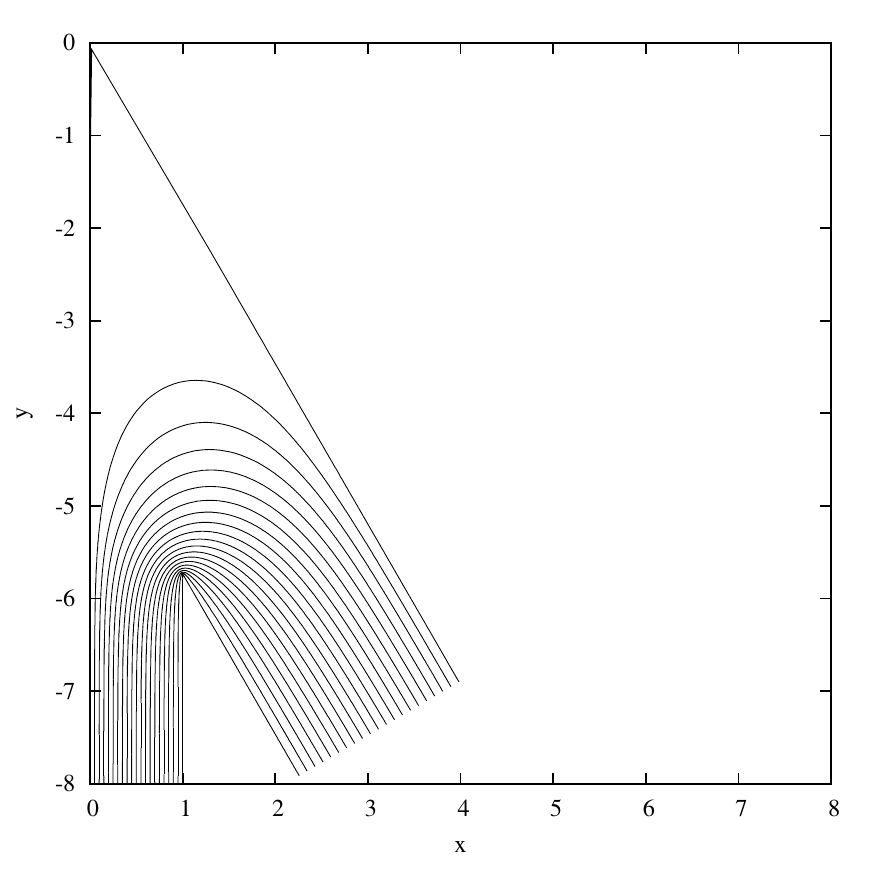}
\caption{Distribution of a current density for  $\alpha=30^{\circ}$ and $h/k=2$.}
\label{30}
\end{figure}

\item $\alpha=120^{\circ}$:

\begin{multline}
z=C\left\{-\frac{1}{b}\ln(t+b)
+\frac{1}{2b}\ln(t^2-bt+b^2)
-\ln(t-1)
+\frac{1}{2}\ln(t^2+t+1)\right.
\\
\left.-\frac{i\sqrt{3}}{2b}\ln\left(\frac{2t-b-i\sqrt{3}b}{-2t+b-i\sqrt{3}b}\right)
+\frac{i\sqrt{3}}{2}\ln\left(\frac{2t+1-i\sqrt{3}}{-2t-1-i\sqrt{3}}\right)\right\}+C_1
\label{Expr120}
\end{multline}
where
\begin{equation*}
C=\frac{h}{2\pi}(\sqrt{3}+i)
\end{equation*}
\begin{equation*}
C_1=\frac{h-k}{4}+i\frac{\sqrt{3}}{12}(7k+h)
\end{equation*}
\begin{equation*}
t=\left(\frac{z_1+b^3}{z_1-1}\right)^{\tfrac{1}{3}}
\end{equation*}
\begin{equation*}
b=\frac{h}{k}.
\end{equation*}
\end{enumerate}
The lines of current calculated for these angles are shown in figures~\ref{60} -- \ref{120}.

The current strength $J$ applied to the conductor supposed to be specified. The current distribution for $y\to -\infty$ (fig.~\ref{60}--\ref{120}) is uniform therefore one can assume that current density at an infinite distance is $j_{-\infty}=J/k$. On the other hand, current density is proportional to the amplitude of complex potential $W(z_1)$ derivative \cite{Smajt}:
\begin{displaymath}
j(z)=A'\left|\frac{\partial W(z_1(z))}{\partial z}\right|.
\end{displaymath}
The complex potential of a point charge placed at the origin $W(z_1)\sim\ln(z_1)$ \cite{Smajt} therefore one can obtain the expression for the current density in terms of the implicit function $z_1(z)$
\begin{equation}
j(z)=A'\left|\frac{\partial W(z_1)}{\partial z_1}\frac{dz_1}{dz}\right|=
A\left|\frac{1}{C}\left(\frac{z_1(z)+a}{z_1(z)-1}\right)^{1-\beta}\right|
\label{current1}
\end{equation}
where $A'$ and $A$ are proportionality factors.

$y\to -\infty$ and $x=\mathrm{const}$ correspond to value $z_1=\left.r_1 e^{\theta_1}\right|_{r_1\to 0,\:\theta_1\in[0,\pi]}$. The substitution of these values into (\ref{current1}) gives
\begin{equation}
j_{-\infty}=A\left|\frac{\pi}{h}ib\right|=A\frac{\pi}{k}
\label{j_infty}
\end{equation}
whence it follows that
\begin{equation*}
A=\frac{J}{\pi}.
\end{equation*}

Finally, the distribution of current density is the following
\begin{equation}
j(z)=\frac{J}{\pi}\left|\frac{1}{C}\left(\frac{z_1(z)+a}{z_1(z)-1}\right)^{1-\beta}\right|.
\label{current}
\end{equation}

\begin{figure}[!t]
\centering
\includegraphics[width=3.4in]{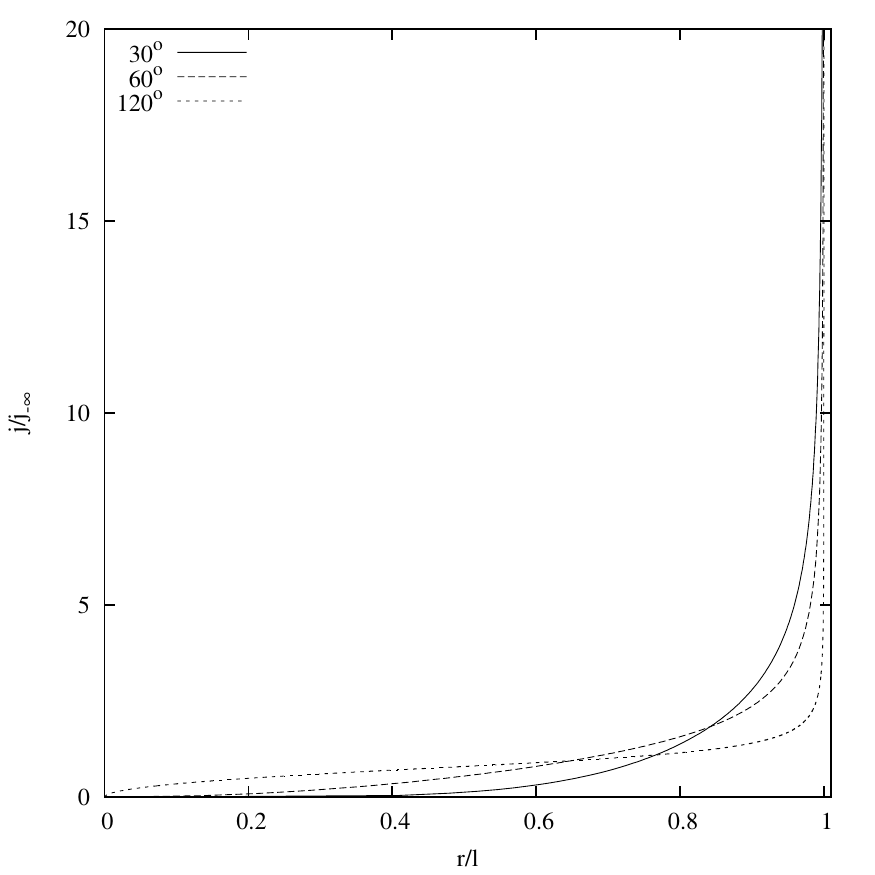}
\caption{The dependency between scaled current density and scaled distance from the point $B$ along the line $BE$ (fig.~\ref{post_zadachi}) for the angles of 30$^{\circ}$, 60$^{\circ}$ and 120$^{\circ}$. $l$ is the length of $BE$.}
\label{general_result}
\end{figure}

The dependency $z_1(z)$ is given by the expression (\ref{Int}) (or (\ref{Expr60}) - (\ref{Expr120})) in implicit form and can be calculated numerically. The dependency between current density and the distance from the point $B$ along the line $BE$ (fig.~\ref{post_zadachi}) is depicted in figure \ref{general_result}. The scaling is used in order to compare distributions for different angles.

The point $E$ of the complex plane $z$ corresponds $z_1=1+i\cdot 0$ (fig.~\ref{post_zadachi}). With relation to (\ref{current}) one can see that the current density in the angle is infinite. This result is nonphysical because in real wires there aren't absolutely sharp angles. 

\section{Current distributions in wires with rounded angles}

To take the rounding into account we have used a technique similar to the one considered in \cite{Lavrentiev} but we have used three new parameters instead of two. It was done in order to obtain smooths curve that looks like an arc of a circle.

\begin{figure}[!t]
\centering
\includegraphics[width=3.4in]{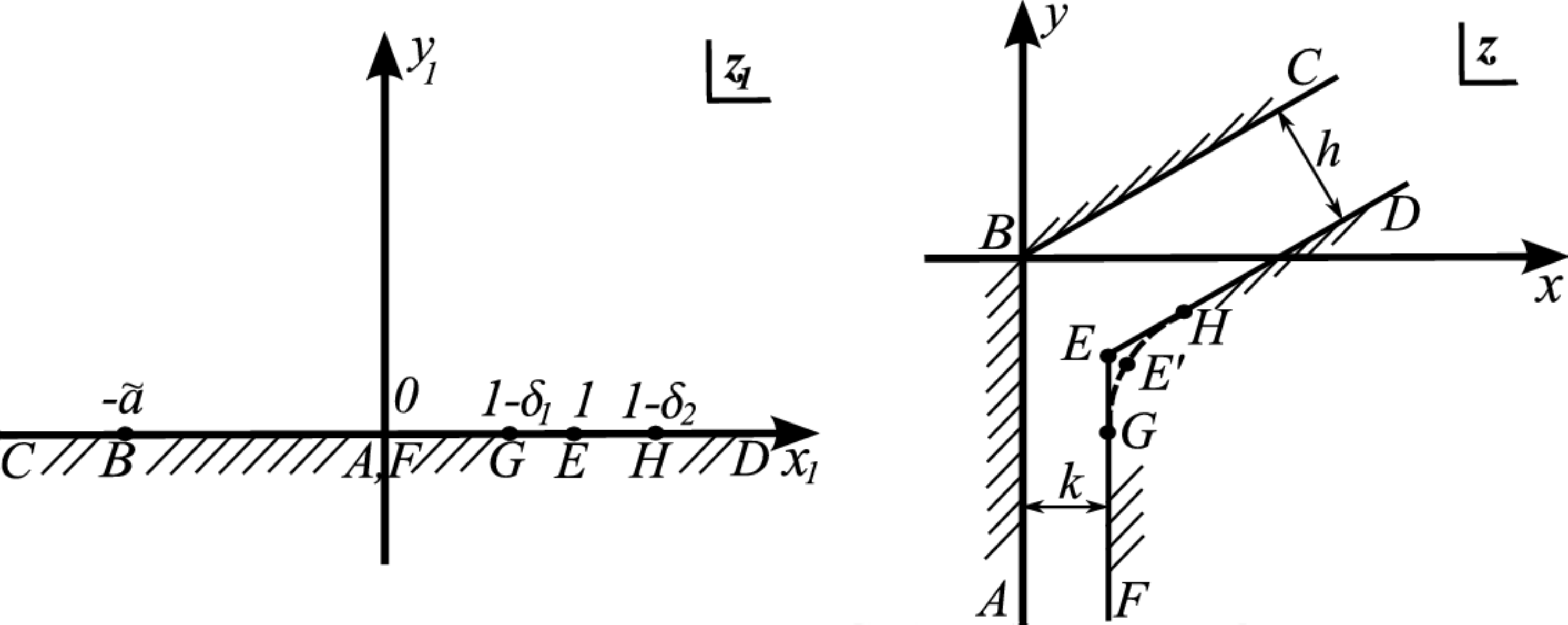}
\caption{The points conformity for a conductor with the rounded angle.}
\label{rounding}
\end{figure}

We have modified the expression (\ref{Schwarz-Cristoffel}) in the following way
\begin{multline}
z_R=\tilde{C}\int\frac{1}{z_1}\left(\frac{z_1-1}{z_1+\tilde{a}}\right)^{1-\beta}dz_1\\
+\tilde{C}\gamma\int\frac{1}{z_1}\left(\frac{z_1-1+\delta_1}{z_1+\tilde{a}}\right)^{1-\beta}dz_1
+\tilde{C}\gamma\int\frac{1}{z_1}\left(\frac{z_1-1-\delta_2}{z_1+\tilde{a}}\right)^{1-\beta}dz_1
\label{RoundSC}
\end{multline}
where constants $\delta_1$, $\delta_2$ and $\gamma$ define the radius of rounding (fig.~\ref{rounding}).

Constants $\tilde{a}$ and $\tilde{C}$ were defined in a similar way to the non-rounded case from the conditions of boundary conformity
\begin{gather*}
\tilde{C}=\frac{h}{\pi(1+2\gamma)}(\sin\alpha-i\cos\alpha)\\
\tilde{a}=\left(\frac{h}{k}\right)^{\tfrac{1}{1-\beta}}
\left\{\frac{1+\gamma(1-\delta_1)^{1-\beta}+\gamma(1-\delta_2)^{1-\beta}}{1+2\gamma}\right\}^{\tfrac{1}{1-\beta}}.
\end{gather*}

To calculate the expression (\ref{RoundSC}) we have supposed again that $1-\beta=P/Q$ and denoted 
\begin{displaymath}
z_R=I(1)+\gamma I(1-\delta_1)+\gamma I(1+\delta_2)
\end{displaymath}
where
\begin{displaymath}
I(m)\equiv\tilde{C}\int\frac{1}{z_1}\left(\frac{z_1-m}{z_1+\tilde{a}}\right)^{\tfrac{P}{Q}}dz_1.
\end{displaymath}
Changing variables
\begin{gather*}
t_m=\left(\frac{z_1+\tilde{a}}{z_1-m}\right)^{\tfrac{1}{Q}}\\
b_m=\left(\frac{\tilde{a}}{m}\right)^{\tfrac{1}{Q}}
\end{gather*}
we have led $I(m)$ to the rational integral
\begin{equation}
I(m)=-(1+b_m^Q)Q\tilde{C}\int\frac{t_m^{Q-1}}{t_m^P(t_m^Q+b_m^Q)(t_m^Q-1)}dt.
\label{RoundInt}
\end{equation}

The obtained expression is formally equivalent to (\ref{Int}) therefore previously calculated expressions (\ref{Expr60}) -- (\ref{Expr120}) have been used to find the result.

\begin{figure}[!t]
\centering
\includegraphics[width=2.0in]{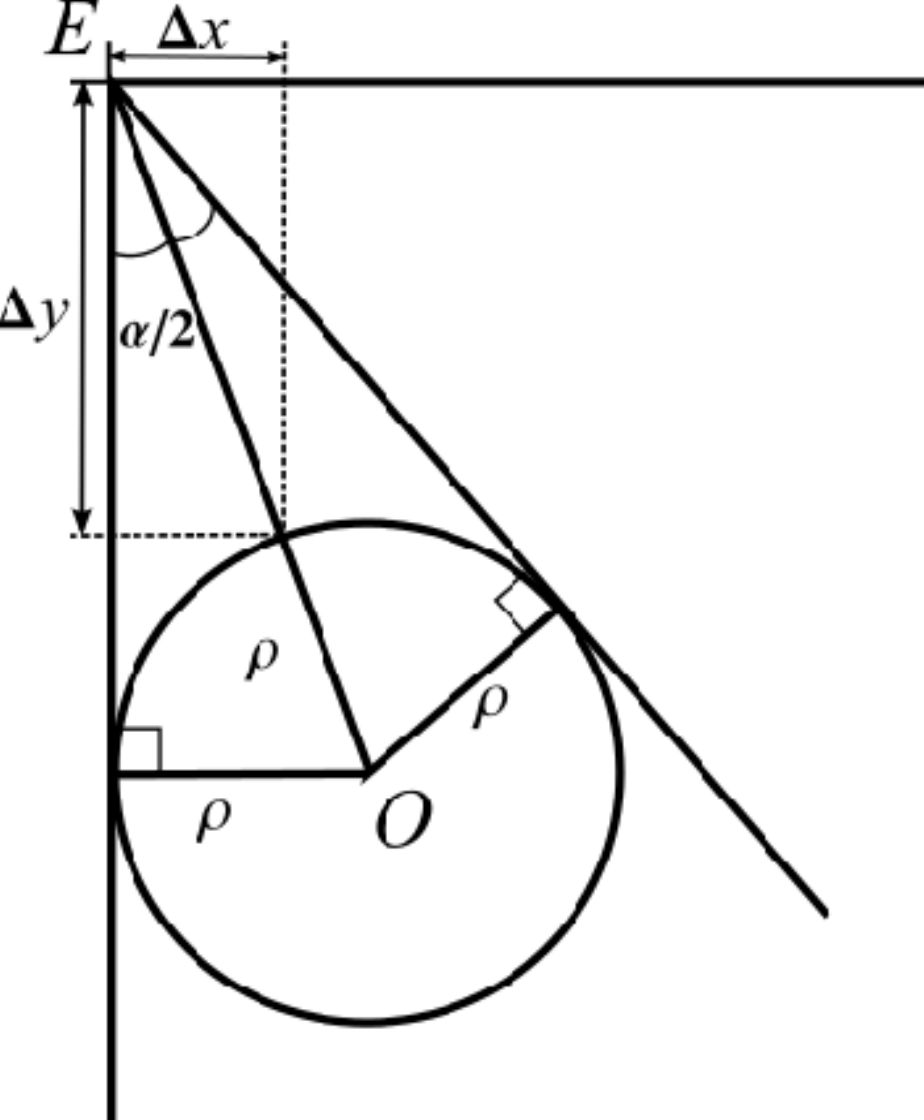}
\caption{Schematic diagram of a rounded angle.}
\label{geometry}
\end{figure}

To determine the dependency between the radius of corner rounding $\rho$ and parameters $\delta_1$, $\delta_2$ and $\gamma$ we have assumed that the rounding curve is an arc of a circle (fig.~\ref{geometry}). This assumption is correct for $\rho\ll h,k$. The placement of the point $E$ corresponds to $z(1)$ and the placement of the point $E'$ corresponds to $z_{R}(1)$. The lengths $\Delta x$ and $\Delta y$ were expressed through the radius $\rho$ using simple geometric considerations
\begin{gather}
\Delta x=\mathrm{Re}[z_R(1)]-x_E\cong\rho\left(1-\sin\frac{\alpha}{2}\right)
\label{Dx}\\
\Delta y=y_E-\mathrm{Im}[z_R(1)]\cong\rho\left(1-\sin\frac{\alpha}{2}\right)\cot\frac{\alpha}{2}.
\label{Dy}
\end{gather}
The equivalence is inaccurate because the real curve isn't an arc of a circle. The coordinates of the point $E$ are the following
\begin{gather}
x_E= k
\label{xE}\\
y_E=-\frac{h+k\cos\alpha}{\sin\alpha}.
\label{yE}
\end{gather}

The formulas (\ref{Dx}) -- (\ref{yE}) give two equations for three parameters. To obtain the third equation we have imposed the condition $GE=EH$ (fig.~\ref{rounding}). Finally, the system of equation for parameters $\delta_1$, $\delta_2$ and $\gamma$ has been obtained
\begin{equation}
\left\{
\begin{array}{lcl}
\mathrm{Re}[z_R(1)]\cong k+\rho\left(1-\sin\dfrac{\alpha}{2}\right)\\
\mathrm{Im}[z_R(1)]\cong-\dfrac{h+k\cos\alpha}{\sin\alpha}-\rho\left(1-\sin\dfrac{\alpha}{2}\right)\cot\dfrac{\alpha}{2}\\
\\
|z_R(1)-z_R(1-\delta_1)|^2=|z_R(1)-z_R(1+\delta_2)|^2.
\end{array}
\right.
\label{System_param}
\end{equation}

The expression for current density for such mapping is defined in a similar way to (\ref{current}) and has the following view
\begin{equation*}
j(z_R)=\left|\frac{1}{C}\frac{[z_1+a]^{1-\beta}}{[z_1-1]^{1-\beta}
+\gamma[z_1-1+\delta_1]^{1-\beta}+\gamma[z_1-1-\delta_2]^{1-\beta}}\right|
\end{equation*}
where $z_1=z_1(z_R)$.

The system (\ref{System_param}) was solved numerically for angles of 120$^{\circ}$, 60$^{\circ}$ and 30$^{\circ}$. It was determined that the solution of it exists only for the point $E'$ placed sufficiently close to the point $E$. That imposes the restriction on the value of $\rho$ for every given angle. For instance, acceptable radii for the angle of 120$^{\circ}$ should be less than $0.8k$ for the considered case. For the angle of 30 $^{\circ}$ the maximal radius is just $0.012k$. Probably, bigger radii should be obtained using some other techniques. More detail approach to this problem is the object of further investigation.

\begin{figure}[!t]
\includegraphics[width=3.4in]{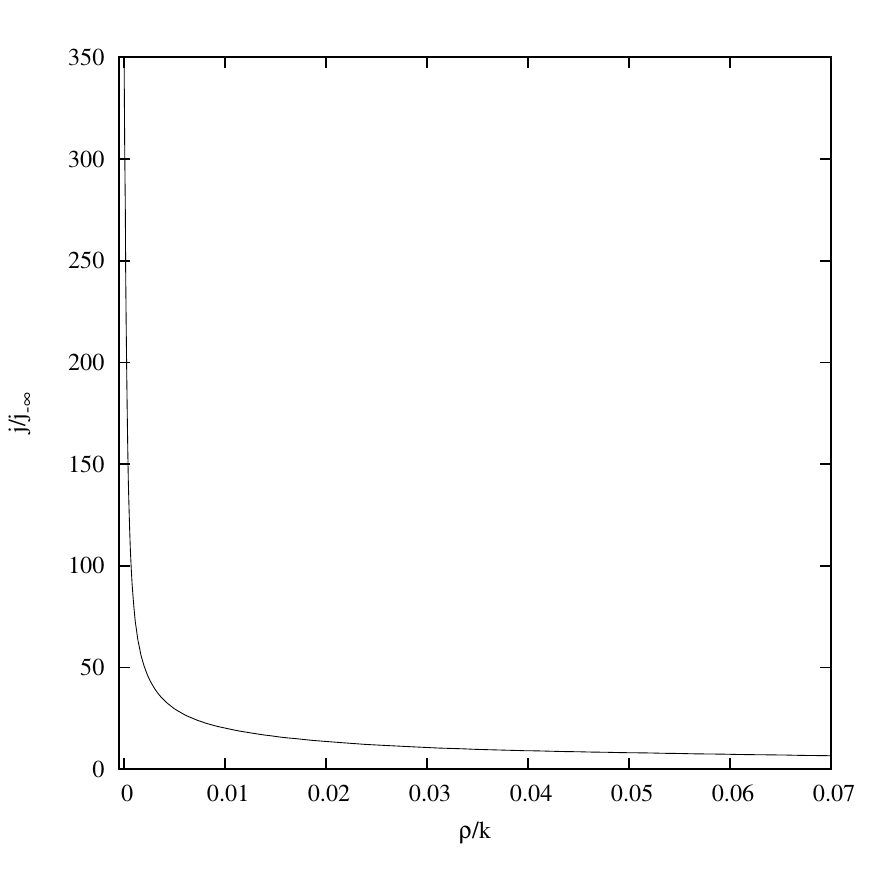}
\caption{Dependency between the density of current at the point $E'$ and the rounding radius $\rho$ for the angle of 60$^{\circ}$.}
\label{rJ60}
\includegraphics[width=3.4in]{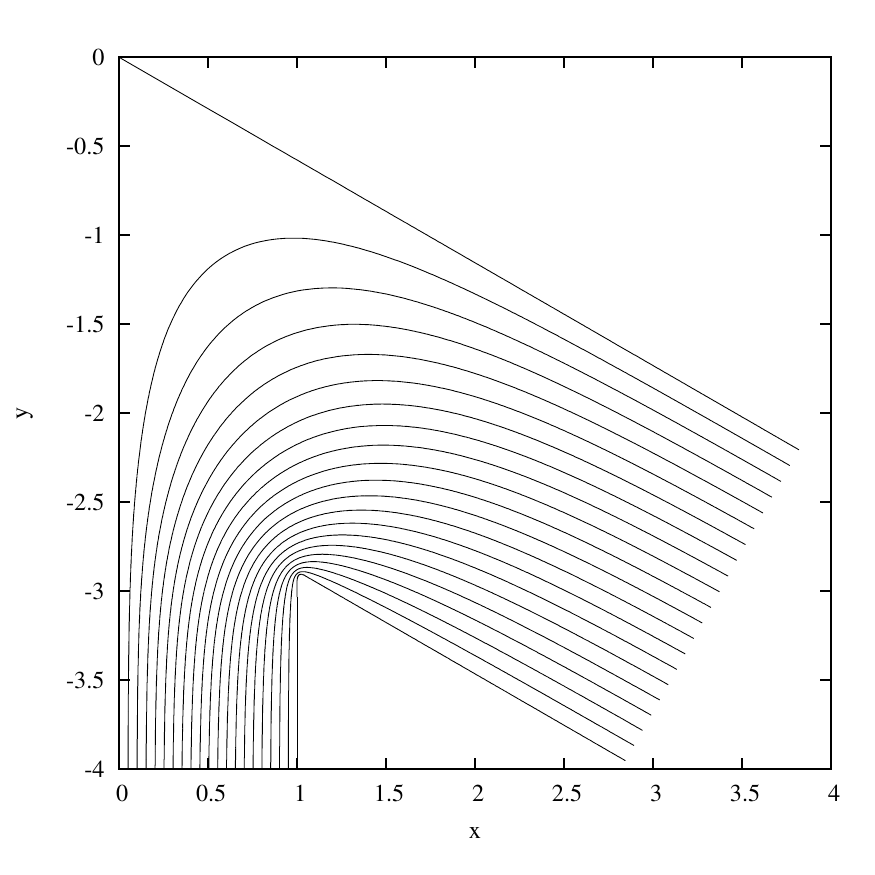}
\caption{Distribution of current in the conductor with the rounded angle of $60^{\circ}$ and $h/k=2$. The rounding radius $\rho=0.02k$.}
\label{60r}
\end{figure}

\begin{figure}[!t]
\centering
\includegraphics[width=3.4in]{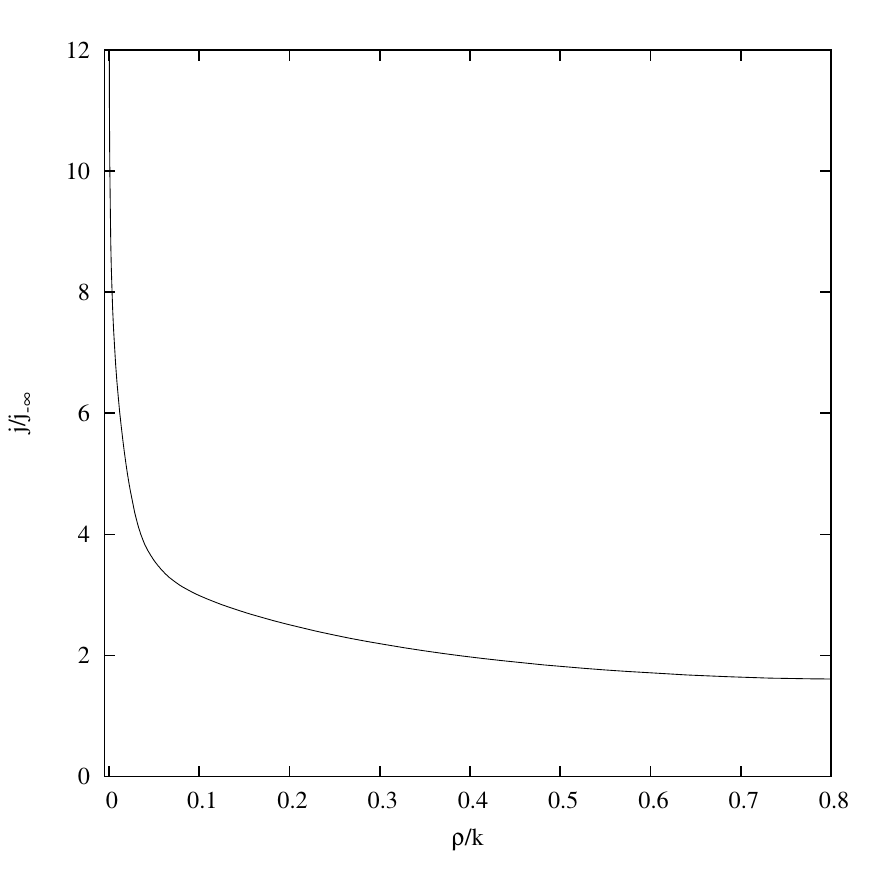}
\caption{Dependency between the density of current at the point $E'$ and the rounding radius $\rho$ for the angle of 120$^{\circ}$.}
\label{rJ120}
\centering
\includegraphics[width=3.4in]{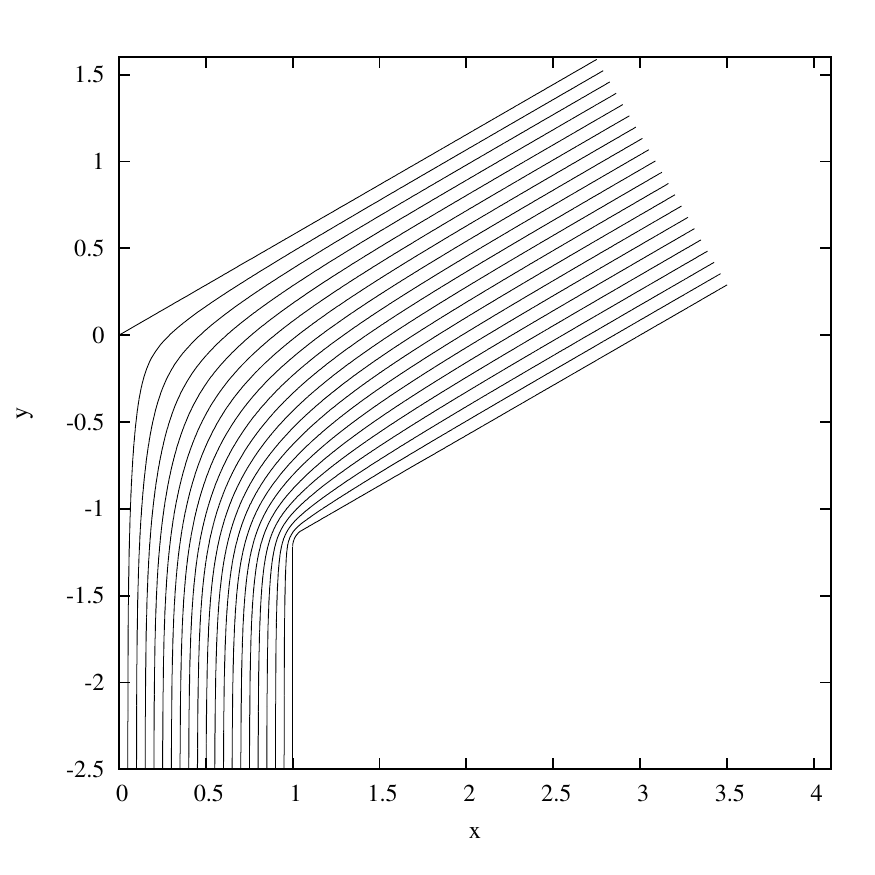}
\caption{Distribution of current in the conductor with the rounded angle of $120^{\circ}$ and $h/k=1.5$. The rounding radius $\rho=0.1k$.}
\label{120r}
\end{figure}

The dependencies between the density of current at the point $E'$ normalised to $j_{-\infty}$ (\ref{j_infty}) and the radius $\rho$ of angle's rounding normalized to the value of $k$ are shown in figures~\ref{rJ60}, \ref{30} and \ref{rJ120}.

\begin{figure}[!t]
\centering
\includegraphics[width=3.4in]{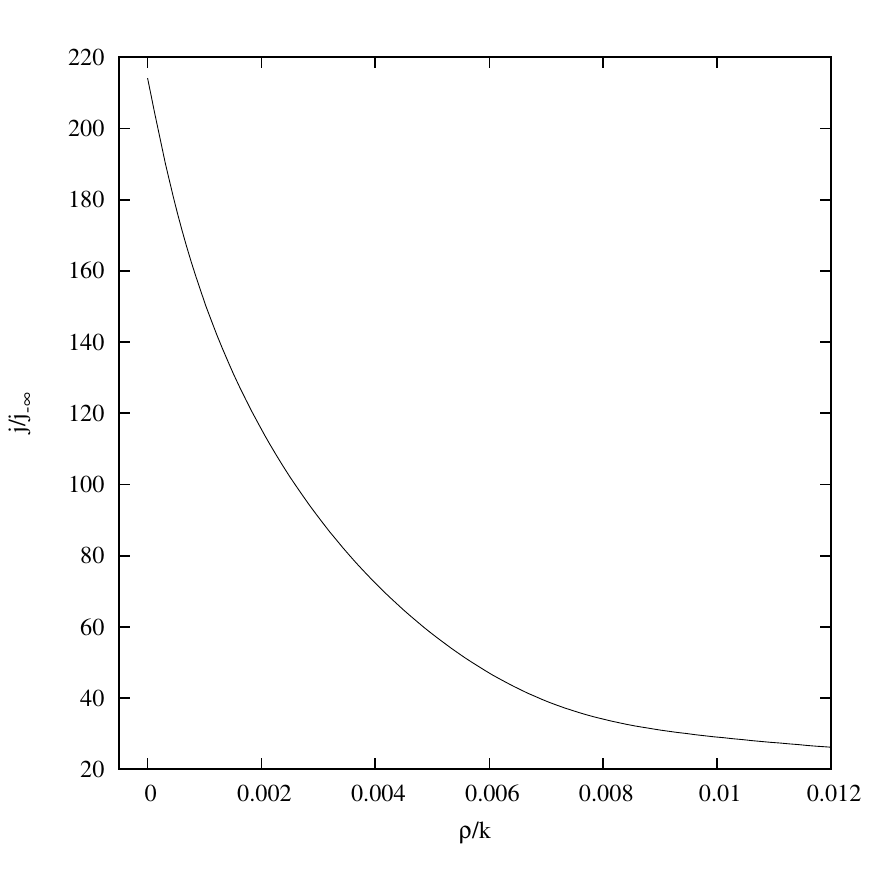}
\caption{Dependency between the density of current at the point $E'$ and the rounding radius $\rho$ for the angle of 30$^{\circ}$.}
\label{rJ30}
\includegraphics[width=3.4in]{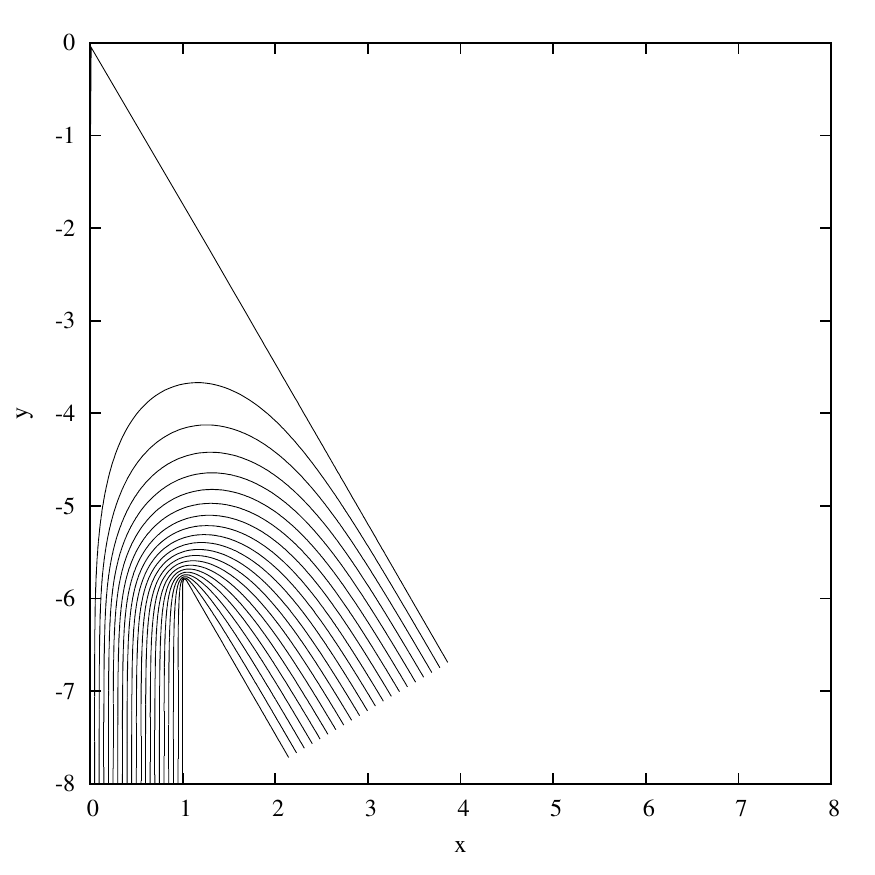}
\caption{Distribution of current in the conductor with the rounded angle of $30^{\circ}$ and $h/k=2$. The rounding radius $\rho=0.011k$.}
\label{30r}
\end{figure}

The current distributions in the conductors with the rounded angles are presented in figures~\ref{60r}, \ref{30r} and \ref{120r}.

\clearpage
\section*{Conclusion}
The obtained results show that the consideration of current distribution in a thin film wires with absolutely sharp angles gives a nonphysical results, namely, an infinite current density in the angle. Therefore the technique of angle's rounding  was suggested. Using of three parameters defining the rounding radius instead of two enables us to obtain a smooth line having a similar form with an arc of a circle. The obtained dependency between the extreme current density in a wire and a rounding radius can be used to estimate a radius of an angle's rounding required to prevent a destruction of a wire with an applied current.

\bibliographystyle{plain}
\bibliography{article}

\end{document}